\documentclass{aa}
\usepackage{graphicx}
\begin{document}
   \title{Systematic Construction of Exact 2-D MHD Equilibria \\
          with Steady, Compressible Flow in Cartesian Geometry and Uniform Gravity}

   \subtitle{}

   \titlerunning{Systematic Exact MHD Models}

   \author{G.J.D. Petrie
          \inst{1}\fnmsep\thanks{Present address:
             Section of Astrophysics, Astronomy and Mechanics, 
             Department of Physics, University of Athens,
             Panepistimiopolis, GR-157 84 Zografos,
             Athens, Greece }
          \and
          N. Vlahakis
          \inst{2}
          \and
          K. Tsinganos
             \inst{3}\fnmsep\thanks{Present address: As GP}
          }

   \offprints{G.J.D. Petrie}

   \institute{Department of Physics, University of Crete, PO Box 2208,
              GR-710 03 Heraklion, Crete, Greece\\
              \email{gordonp@physics.uoc.gr}
         \and
              Department of Astronomy \& Astrophysics and 
              Enrico Fermi Institute, University of Chicago,
              5640 S. Ellis Ave., Chicago, IL 60637, USA\\
              \email{vlahakis@jets.uchicago.edu}
         \and
             Department of Physics, University of Crete, PO Box 2208,
              GR-710 03 Heraklion, Crete, Greece\\
             \email{tsingan@physics.uoc.gr}
             }

   \date{Received;accepted}

   \abstract{We present a systematic method for constructing two-dimensional magnetohydrodynamic equilibria with compressible flow in Cartesian geometry.  This systematic method has already been developed in spherical geometry and applied in modelling solar and stellar winds and outflows (Vlahakis \& Tsinganos,1998) but is derived here in Cartesian geometry in the context of the solar atmosphere for the first time.  Using the method we find several new classes of solutions, some of which generalise known solutions, including the Kippenhahn \& Schl\"uter (1957) and Hood \& Anzer (1990) solar prominence models and the Tsinganos, Surlantzis \& Priest (1993) coronal loop model with flow, and some of which are completely new.  Having developed the method in full and summarised the several classes of solutions, we explore in a some detail one of the classes to illustrate the general construction method.  From one of the new classes of solutions we calculate two loop-like solutions, one of which is the first exact two-dimensional magnetohydrodynamic equilibrium with trans-Alfv\'enic flow.
   \keywords{MHD --  Methods: analytical -- Sun: corona -- Sun: magnetic fields}
   }

   \maketitle
%

\section{Introduction}
\label{introduction}

The investigation of plasma equilibria is one of the most important
problems in magnetohydrodynamics (MHD) and arises in a number of
fields including thermonuclear fusion, astrophysics and solar physics.
At present there are many difficulties surrounding the description of
fully three-dimensional configurations and so it is necessary to
consider configurations with additional symmetry from a mathematical
point of view (but see Petrie \& Neukirch, 1999).
In many astrophysical situations, e.g. solar or stellar winds and
outflows, axial symmetry is important while in solar physics
translational symmetry is common in models of arcades,
loops and prominences.

Even with the steady-flow and symmetry assumptions,
the prospect of solving the full MHD equations is not good
without further simplification.  Therefore most modellers of MHD structures with flow adopt a numerical approach (e.g. Karpen et al., 2001).  However we choose to use analytical techniques as far as possible because of their greater computational economy and ease of use and the smaller role played by boundary conditions in defining a solution as opposed to numerical techniques, as well as the comprehensiveness and clarity with which analytical methods can describe solution spaces.  In this work we impose on our
solutions spatial self-similarity i.e. the assumption of a scaling
law of one of the variables on one of the coordinates.
With this assumption the full MHD equations can be reduced to a
system of ordinary differential equations (ODE) which can be integrated
by standard methods. Distinct from temporally self-similar solutions
such as those by Neukirch \& Cheung (2001) and the steady solutions obtained via a transformation method by Imai (1960) and Webb et al. (1994), many known two-dimensional
equilibria are spatially self-similar both in spherical geometry
(Sauty \& Tsinganos, 1994; Vlahakis \& Tsinganos 1998, 1999, Vlahakis et al. 2000)
and in Cartesian geometry (de Ville \& Priest, 1991;
Tsinganos, Surlantzis \& Priest, 1993; Del Zanna \& Hood, 1996;
Del Zanna \& Chiuderi, 1996).
Two-dimensional equilibria with flow have been used in the context
of the solar atmosphere to model coronal loops and arcades
(de Ville \& Priest, 1991; Tsinganos, Surlantzis \& Priest, 1993)
and prominences (Del Zanna \& Hood, 1996).

A significant proportion of the energy emission from the solar corona
is concentrated along well-defined curved paths called loops (Bray et al., 1991).
Coronal loops are a feature of active regions and it is believed that
they spread themselves out to dominate the lower corona,
particularly in and over active regions.
The loops are believed to trace closed lines of force of
the magnetic field which penetrate the photosphere from below and
expand to fill the whole of the coronal volume above an active region.
Two patterns of steady plasma flow are possible in a loop:
a flow down both legs starting at the apex, and a flow up one
leg and down the other. Upward flows in both legs of loops associated with flares have also been observed but these are time-dependent.  It has been noted that unidirectional
loop flows are characteristic of new and complex active regions
(Bray et al., 1991).
The motion along a loop may continue for up to several hours so that a steady-flow description is valid.

In this paper we present several new classes of two-dimensional MHD
equilibria in a uniform gravitational field, with flow in Cartesian geometry.  
The flow may be strictly sub-Alfv\'enic, super-Alfv\'enic or trans-Alfv\'enic.  In the trans-Alfv\'enic case the solution can be made to cross the Alfv\'en critical point smoothly.  Discontinuities are possible in the solution but we choose to avoid them to demonstrate that this is possible.  We have no observed trans-Alfv\'enic loops to compare with and shocks may occur at critical points in reality but our solution method allows smooth as well as shocked solutions.
The solutions are calculated using a systematic nonlinear separation
of variables construction method already seen in spherical geometry
(Vlahakis \& Tsinganos, 1998) and developed for the first time
in Cartesian geometry in this paper.
The resulting classes of solutions are found to include as special cases
several well-known two-dimensional equilibria as well as many hitherto
unknown solutions.
To illustrate the general construction method we present two example
solutions of a loop-like structure with unidirectional flow.
We present a sub-Alfv\'enic case and a trans-Alfv\'enic case.

The paper is organised as follows.
We introduce the MHD equations briefly and give details of our
assumptions in Sections \ref{MHD} and \ref{assumptions} before using
a simple theorem to construct several classes of translationally
self-similar solutions in Section \ref{construct}, summarised
in Table \ref{largetable}.
In Section \ref{newclass} a special case of one of the classes
is explored in a little detail by way of illustrating the
general solution method and providing for the first time an
exact loop-like solution with trans-Alfv\'enic flow as well as
a new sub-Alfv\'enic example.
The paper is summarised in Section \ref{conclusions}.

\section{Construction of the model}

In this section we describe in some detail how our model can be 
systematically obtained from the closed set of governing full 
MHD equations.

\subsection{Governing equations}
\label{MHD}

The {\it dynamics} of flows in solar coronal loops may be described to zeroth 
order by the well known set of the steady (\(\partial/\partial t=0\))   
ideal hydromagnetic equations: 
\begin{equation}\label{momentum}
\rho \left( \vec{V}\cdot\vec{\nabla}\right)\vec{V}=
\frac{1}{4 \pi} {\left(\vec{\nabla}\times
\vec{B}\right)\times\vec{B} } 
-\vec{\nabla}P-\rho g \hat Z 
\,,
\end{equation}
\begin{equation}\label{fluxes}
\vec{\nabla}\cdot\vec{B}=0\,,\quad 
\vec{\nabla}\cdot\left(\rho\vec{V}\right)=0
\,,\quad \vec{\nabla}\times\left(\vec{V}\times\vec{B}\right)=0
\,,
\end{equation}

\noindent
where \(\vec{B}\), \(\vec{V}\), \(-g \hat Z \)  denote the magnetic,
velocity and (uniform) external gravity fields while \( \rho\) and $P$ the gas density and 
pressure.\\
The {\it energetics} of the outflow on the other hand is governed by  
the first law of thermodynamics : 
\begin{equation}\label{firstlaw}
q=\rho \vec{V} \cdot \left [ \vec{\nabla}\left(\frac{1}{\Gamma-1}
\frac{P}{\rho}\right)+P \vec{\nabla} \frac{1}{\rho} \right ]
\,,
\end{equation}
where $q$ is the net volumetric rate of some energy input/output 
(Low \& Tsinganos 1986), while 
$\Gamma=c_{p}/c_{v}$ with $c_{p}$ and $c_{v}$ the specific heats for an 
ideal gas.\\

With translational Y-symmetry in Cartesian coordinates $(Z, X, Y)$, the 
coordinate $Y$ is ignorable (\(\partial/\partial Y=0\)) and   
we may introduce the poloidal magnetic flux function
(per unit length in the $\hat{Y}$ direction)

\begin{equation}
{\vec B}=\vec \nabla A(Z,X)\times\hat{Y} + B_y (Z,X) \hat Y ,
\end{equation}

\noindent such that several free integrals of $A$ exist. 
They are the ratio of the mass and magnetic fluxes on the poloidal plane (Z-X), $\Psi_A(A)$,

\begin{equation}
{\vec V_p}={{\Psi}_A\over 4\pi\rho}{\vec B_p}
\,,
\end{equation}
where the stream function $\Psi $ is a function of the magnetic flux
function $A$ and ${\Psi}_A$ is its derivative,
the induction potential $\Phi (A)$, $\vec V \times \vec B = \vec \nabla \Phi$, and the 
function $L(A)$ in terms of which the field components in the Y-direction are related as 
$\displaystyle V_y-{B_y \over {\Psi}_A}=L(A)$,
and $\displaystyle V_y-{{\Psi}_A\over 4\pi\rho}B_y=\Phi_A(A)$, 
(Tsinganos 1982), or,
\begin{equation}
\label{ycomponents}
{B_y}=\Psi_A {\Phi_A -L\over 1 - \Psi_A^2/4\pi \rho }\;\;\;\,, 
{V_y}={\Phi_A -L\Psi_A^2/4\pi \rho  \over 1 - \Psi_A^2/4\pi \rho }
\,,\label{ycpts}
\end{equation}
Then, the system of Eqs. (\ref{momentum}) - (\ref{fluxes}) 
reduces to a set of two partial and nonlinear differential equations, 
i.e., the $\hat Z$- and $\hat X$- components of the momentum equation on the 
poloidal plane,
\begin{equation}\label{momentum2}
\frac{1}{4 \pi} {\left( \vec{B}\cdot\vec{\nabla}\right)(1-\Psi_A^2/4\pi\rho)\vec{B} } =  
\vec{\nabla}\left ( P+{B^2\over 8\pi} \right )  + \rho g \hat Z 
\,.
\end{equation}

 
The system of equations (\ref{momentum}) - (\ref{fluxes}) - (\ref{firstlaw})  
can be solved to give $\rho$, $P$,  
$\vec V$ and $\vec B$ \underbar{if} the heating function $q$ 
is known. 
Similarly, one may close this system of Eqs.   
(\ref{momentum}) - (\ref{fluxes}) - (\ref{firstlaw}), \underbar{if} an 
extra functional relationship of $q$ with the unknowns $\rho$, $P$ and $A$ exists. 
As an example, consider the following {\it special} functional relation of $q$ 
with these unknowns $\rho$, $P$ and $A$ (Tsinganos, Trussoni \& Sauty 1992),
\begin{equation}\label{polytropic_heating}
q=\frac{\gamma-\Gamma}{\Gamma-1}
\frac{P}{\rho}\vec{V} \cdot \vec{\nabla}\rho
\,,
\end{equation}
where $\gamma \le \Gamma$=5/3. 
Then, Eq.(\ref{firstlaw}) can be integrated at once to give the familiar 
polytropic relation between $P$ and $\rho$, 
\begin{equation}\label{polytropic}
P=Q \left( A \right)\rho ^{\gamma}
\,,
\end{equation}
for some function $Q (A)$ corresponding to the enthalpy along a surface $A=const$. 
In this special case we can integrate the projection of the momentum 
equation along a stream-field line $A=const$ on the poloidal plane
to get the well known Bernoulli integral
which subsequently can be combined with the component of the momentum 
equation across the poloidal field lines (the transfield equation) to yield 
$\rho$ and $A$.
After finding a solution, one may go back to Eq.(\ref{firstlaw}) and fully 
determine the function $q(Z, X )$. 
It is evident that even in this {\it special} polytropic case 
with $\gamma \neq \Gamma$ the heating function $q$ (not its functional form 
but the function $q(Z, X )$ itself) can be found only {\it a posteriori}.
Note that for $\gamma=\Gamma$ and only then the flow is isentropic. 
\\
Evidently, it is not possible to integrate Eq.(\ref{firstlaw}) for {\it any}  
functional form of the heating function $q$, such as it was possible with 
the special form of the heating function given in 
Eq.(\ref{polytropic_heating}).
To proceed further then and find other more general solutions (effectively having a variable value for $\gamma= {\partial \ln P(\rho\,,A)}/{\partial \ln \rho} $ as $P$ is no longer integrable as a power of $\rho$), 
one may choose some other functional form for the heating function $q$ and 
from energy conservation, Eq.(\ref{firstlaw}), derive a functional form 
for the pressure.
Equivalently, one may choose a functional form for the pressure $P$ and 
determine the volumetric rate of thermal energy {\it a posteriori}  
from Eq.(\ref{firstlaw}), after finding the expressions of 
$\rho$, $P$ and $A$ which satisfy the two remaining components of the 
momentum equation. 
Hence, in such a treatment the heating sources which produce some specific 
solution are not known {\it a priori}; instead, they can be determined 
only {\it a posteriori}.
However, it is worth keeping in mind that as explained before, this situation is 
analogous to the more familiar constant $\gamma$ polytropic case, with $\gamma 
\neq \Gamma $. In this paper we shall follow this approach, which is further 
illustrated in the following section.

However, even by adopting this approach, the integration of the system 
of mixed 
elliptic/hyperbolic partial differential equations (\ref{momentum}) - 
(\ref{fluxes}) remains a nontrivial undertaking. Besides their 
nonlinearity, the difficulty is largely due to the fact that a physically 
interesting solution is constrained to cross some critical surfaces which 
are not known {\it a priori} 
but are determined simultaneously with the solution. 

\subsection{Assumptions}
\label{assumptions}

In order to construct analytically some classes of exact solutions for coronal 
loops, we shall first assume in the case of solutions with trans-Alfv\'enic flows that the magnetic loop is planar, $B_y=0$,   
and there exists only a velocity component along the y-direction 
which is constant on each poloidal field line, $V_y = \Phi_A (A) = L(A)$.  These assumptions are necessary for a trans-Alfv\'enic solution to be finite at an Alfv\'en point $M^2=1$ (see Eqs. (\ref{ycpts})).
Note that with the assumption of a planar magnetic field the function $V_y = \Phi_A (A)=L(A)$ is completely
decoupled from the equations describing the flow and is simply given in 
terms of the free integral $\Phi_A (A)$, or, $L(A)$.  A sub-Alfv\'enic or super-Alfv\'enic solution may have a non-planar magnetic field component as well as a non-planar velocity field component and be finite everywhere, provided that Eqs. (\ref{ycpts}) are satisfied, where $\Phi_A (A)$ and $L(A)$ are free integrals not necessarily equal in solutions where Alfv\'en points are absent.  In the present paper we will briefly discuss the possible profiles for $B_y$ and $V_y$ for our simple examples.  In the meantime we develop loop-like solutions by concentrating on the problem in the $x$-$z$ plane.
To  proceed further we shall make the following two key assumptions:
 \begin{enumerate}
  \item that the Alfv\'en number $M$ is solely a function of the
dimensionless horizontal distance $x=X/ Z_0$, i.e.,

\begin{equation}\label{Alfven}
M_{}^{2}=\frac{4 \pi\rho V_{p}^{2}}{B_{p}^{2}}=\frac {\Psi_{A}^{2}}{4 \pi 
\rho} = M^2(x)\,, \label{M2}
\end{equation} 
and
  \item that the poloidal velocity and magnetic fields have an exponential 
dependence on $z=Z/Z_0$,
\begin{equation}\label{assumptions2}
A= Z_0 B_{0} {\cal A}\left(\alpha\right)\,,
\qquad
\alpha=G(x) \exp{(-z)}
\,,\label{G}
\end{equation}
\end{enumerate}

\noindent
for some function $G(x)$, where $Z_0$ and $B_0$ are constants.  
With this formulation the magnetic field has the form

\begin{equation}
{\vec B}= B_0 \alpha {\cal A}^{'}(\alpha) \left[\hat{X}+F(x) \hat{Z}\right] 
\,,
\end{equation}
where
\begin{equation}
F(x) = \frac{1}{G(x)}\frac{dG(x)}{dx}  = \left(\frac{dZ}{dX}\right)_A
\end{equation}
is the slope of the field lines.

We expect that assumptions (1) - (2) are physically reasonable 
for describing the properties of loop flows, at least close to the solar surface. 
\\ 
Instead of using the free functions of $\alpha$, 
(${\cal A}\,, \Psi_A$), we found it more convenient to work instead with the 
two dimensionless functions of $\alpha$, ($g_1$\,, $g_2$),

\begin{equation}\label{g1}
g_1\left( \alpha \right)= \int \alpha {\cal A}^{'2} d\alpha\,,
\end{equation}

\begin{equation}\label{g2}
g_2\left(\alpha \right)=\frac{g Z_0}{B_0^2} \int \frac{\Psi_A ^2}{\alpha } d \alpha \,. 
\end{equation}

Also, we shall indicate by ${\Pi}$ the dimensionless total pressure,

\begin{equation} 
\Pi = \frac{8 \pi}{B_{0}^2}  \left(
P+\frac{B^2}{8 \pi}\right)\, . 
\end{equation}

Using the fact that $A$ is constant along field lines we can collect the
inertial term and the magnetic tension

\begin{eqnarray}
{{(\vec B\cdot\vec \nabla )\vec B}\over 4\pi}-\rho {(\vec V \cdot \vec \nabla )\vec V} =
{1\over 4\pi}{(\vec B\cdot\vec \nabla )}(1-M^2){\vec B} = 
\nonumber \\
= \frac{B_0^2}{4\pi Z_0} 
\alpha g_1^{'}
\left\{ -M^{2'}\hat{X}+\left[(1-M^2) F^{'}-F{M^{2'}}\right] \hat{Z} \right\} \,,
\end{eqnarray}
where $M^{2'} = dM^2(x)/dx$ and $F^{'} = dF(x)/dx$. 
The previous momentum-balance equation can be written as the following system for $\Pi$
\begin{equation} 
\frac{1}{2}{\partial \Pi (x,z)\over\partial x} = 
- \alpha g_1^{'} M^{2'} \,,
\end{equation}
\begin{equation} 
\frac{1}{2} {\partial \Pi (x,z)\over\partial z} = 
\alpha g_1^{'} \left[ (1-M^2) F^{'} - F M^{2'} \right] 
-\alpha g_2^{'} \frac{1}{M^2} \,.
\end{equation}

For any differentiable function ${\cal D}(x,z)$ the following relations hold

\begin{eqnarray}
{\partial {\cal D}(x,z)\over\partial x} & = & {\partial {\cal D}(x,\alpha )\over\partial
x}+ F \alpha {\partial {\cal D}(x,\alpha )\over\partial\alpha} \,, \\
{\partial {\cal D}(x,z)\over\partial z} & = & -\alpha{\partial {\cal D}(x,\alpha
)\over\partial\alpha} \,. 
\end{eqnarray}

Hence, the following differential equations are obtained by using the coordinates
$x$ and $\alpha $

\begin{eqnarray}
\frac{1}{2}{\partial \Pi (x,\alpha )\over\partial x} & = & \alpha g_1'(\alpha
) f_4(x) +\alpha g_2'(\alpha )f_5(x) ,\label{alphaeqn}\\
\frac{1}{2}{\partial \Pi (x,\alpha )\over\partial\alpha} & = & g_1'(\alpha) f_2(x) 
+g_2'(\alpha ) f_3(x).\label{xeqn}
\end{eqnarray}
in terms of the functions of $x$

\begin{eqnarray}
f_1(x) & = & {1\over 2}\left( 1+F^2\right) ,\\
f_2(x) & = & FM^{2'}- F^{'} (1-M^2) ,\\
f_3(x) & = & {1\over M^2}\\
f_4(x) & = & -( 1+F^2)M^{2'}+(1-M^2) F F^{'} ,\\
f_5(x) & = & -{F\ \over M^2} .
\end{eqnarray}

We are now in a position to integrate Eq.(\ref{xeqn}) to obtain an
expression for the total pressure

\begin{equation}
\Pi (x,\alpha )= 2[f_2(x)g_1(\alpha )+f_3(x)g_2(\alpha )+f_0(x)],
\end{equation}

\noindent where $f_0(x)$ is arbitrary, and to substitute this into (\ref{alphaeqn}) to
obtain

\begin{equation}
f_0'+g_1f_2'-\alpha g_1'f_4 +g_2f_3'-\alpha g_2'f_5=0
\,.
\label{singleeq}
\end{equation}
The above equation can be put to the form,
\begin{equation}
Y_1(\alpha)X_1(x) + 
Y_2(\alpha)X_2(x) + 
\dots +Y_5(\alpha )X_5 (x) =0
\,,
\end{equation}
where,
\begin{equation}
Y_1= 1\,, Y_2 = g_1 \,, Y_3 = -\alpha g_1^{'} \,, 
Y_4 = g_2 \,, Y_5 = -\alpha g_2^{'} \,,
\label{Ys}
\end{equation}
and
\begin{equation}
X_1= f^{'}_0\,, X_2 = f^{'}_2 \,, X_3 = f_4 \,, 
X_4 = f^{'}_3 \,, X_5 = f_5 
\,.
\label{Xs}
\end{equation}

We have therefore reduced the MHD equations to a single differential equation whose
terms all have separated coordinate dependence.  Note that
the physical quantities can be recovered via

\begin{eqnarray}
A & = & Z_0 B_0 \int\sqrt{g_1'\over\alpha}d\alpha ,\\
\Psi_A & = & \frac{B_0}{\sqrt{gZ_0}} \sqrt{\alpha g_2^{'} } ,\\
\rho & = & \frac{B_0^2}{4 \pi g Z_0} \frac{\alpha g_2^{'}}{M^2} ,\\
P(x,\alpha ) & = & \frac{B_0^2}{4\pi} \left[ f_0-f_1 \alpha g_1^{'}+f_2g_1+f_3g_2\right] ,\label{pressuresum}\\
{\vec B} & =& B_0 \sqrt{\alpha g_1^{'}} 
\left[\hat{X}+F(x) \hat{Z}\right] + B_y \hat{Y} ,\\ 
{\vec V} & =& \sqrt{g Z_0} \sqrt{\frac{g_1^{'}}{g_2^{'}}} M^2
\left[\hat{X}+F(x) \hat{Z}\right] + V_y \hat{Y} .
\end{eqnarray}

Using our assumptions Eqs. (\ref{M2}, \ref{G}) we have simplified the full MHD equations sufficiently to undertake the task of solving them systematically.  The system of Eqs. (\ref{momentum}, \ref{fluxes}) has been reduced to the single Eq. (\ref{singleeq}).  We undertake this task in the next section.

\subsection{Construction of models}
\label{construct}

Eq.(\ref{singleeq}) can be solved with the help of the following theorem (Vlahakis 1998).

\noindent Theorem: Let $F_n(\alpha )$, $Y_i(\alpha )$ and $X_i(x)$, $i=1,2,\dots ,n$
be functions of $\alpha$ and $x$ such that the relation

\begin{equation}
F_n(\alpha )=Y_1(\alpha) X_1(x)+Y_2(\alpha) X_2(x)+\dots
+Y_n(\alpha) X_n(x)
\end{equation}

\noindent holds.  Then there exist constants $c_1, c_2,\dots , c_n$ such that

\begin{equation}
F_n(\alpha )=c_1 Y_1(\alpha )+c_2 Y_2(\alpha )+\dots +c_n Y_n(\alpha ).
\end{equation}

Note that Eq.(\ref{singleeq}) is a relation of the form

\begin{equation}
Y_n(\alpha )X_n(x)+\dots +Y_1(\alpha )X_1(x)=0\,, 
\label{relationtype}
\end{equation}
with n=5. 
Either (i) $X_n(x)=0$ for every $x$ in which case (indicated by the digit
``0'') we have

\begin{equation}
Y_{n-1}(\alpha )X_{n-1}(x)+\dots +Y_1(\alpha )X_1(x)=0
\,,
\end{equation}

or (ii) $X_n(x)\neq 0$ in which case (indicated by the digit ``1'') we
have

\begin{equation}
Y_n(\alpha )=-{X_1(x)\over X_n(x)}Y_1(\alpha )-\dots -{X_{n-1}(x)\over
X_n(x)}Y_{n-1}(\alpha )
\,.\label{initialsum}
\end{equation}

Therefore according to the theorem there exist constants $\mu_i^{(n)}$,
$i=1,\dots , n-1$ such that $Y_n(\alpha )=\displaystyle \sum_{i=1}^{n-1}
\mu_i^{(n)}Y_i(\alpha )$ giving a condition between the functions of
$\alpha $.  Substituting in Eq.(\ref{initialsum}) we obtain

\begin{equation}
\sum_{i=1}^{n-1} [X_i(x)+\mu_i^{(n)}X_n(x)]\ Y_i(\alpha )=0.
\end{equation}

In both cases (i) and (ii) we obtain a sum with $n-1$ terms and we can
continue to apply this algorithm until we have only one term.  Since for
each of the $n$ terms in Eq.(\ref{relationtype}) there are the above two
possibilities we may obtain $2^n$ cases each corresponding to a number
$xx\dots xx$ where each $x$ is either $0$ or $1$.  The number of $1$'s is
the number of conditions between functions of $\alpha $ while the number
of $0$'s is the number of conditions between functions of $x$.

From our relation (\ref{singleeq}) we can obtain $2^5$ solutions, each
corresponding to a number $xxxxx$. However many of them are not of
interest to us at present. In the first application of the algorithm
$X_5=f_5=-F/M^2$ which must be non-zero if we are to have curved
field lines, and so any solution of interest will have the first digit $1$.
Meanwhile $Y_1=1\neq 0$ so that the last digit must be $0$.  Furthermore
any case $1xx10$ has $g_1(\alpha) $ constant giving $\vec B=\vec V=0$.  This
leaves us with the four cases $1xx00$. There are three unknown functions
of $x$: $G(x)$, $M(x)$ and $f_0(x)$, and two unknown functions of $\alpha
$: $g_1(\alpha )$ and $g_2(\alpha )$. The cases $10100$ and $11000$ have
three equations in $x$ and two in $\alpha $ and are well-determined.  The
case $10000$ has four mutually-incompatible $x$-dependent equations and we
must dismiss this case, but in the case $11100$, although the three
equations in $\alpha $ over-determine the $\alpha $-dependent problem,
useful solutions are available. We describe the $\alpha$-dependent
sets for the families $10100$, $11000$ and $11100$ below.

For the case 10100:
\begin{itemize}
\item
The first digit is 1, so we get
$Y_5$ as a linear combination of $Y_1\,,Y_2\,,Y_3\,,Y_4$.
\item
The second digit is 0, so the corresponding equation is one between
functions of $x$.
\item
The third digit is 1, so we get $Y_3$ as a combination of $Y_1\,,Y_2$.
\footnote{Note that when we reach this step, Eq. (\ref{singleeq}) is of the
form ${\cal H}_3(x) Y_3 + {\cal H}_2(x) Y_2 + {\cal H}_1(x) Y_1=0$,
where ${\cal H}_i$ are combinations of $X_i$.}
\item
The two last digits give equations between functions of $x$.
\end{itemize}
We may neglect the $Y_3$ term in the linear combination of $Y_1\,,Y_2\,,Y_3\,,Y_4$
which give $Y_5$, since $Y_3$ itself is a linear combination of $Y_1\,,Y_2$.

Using Eq.(\ref{Ys}) we get the $\alpha$-dependent set for the case 10100:\\
$$\alpha g_2 '=c_1 +c_2 g_1 +c_4 g_2 \quad \mbox{and} \quad \alpha g_1 '=c_5 +c_6 g_1 \,. $$

In a similar way we may find the $\alpha$-dependent sets for the cases 11000 and 11100.

We summarize the systems of ordinary differential equations
- with respect to $\alpha$ - associated with the three families below.

\begin{equation}
\begin{array}{cl}
{\rm Family} & \qquad \alpha{\rm -dependent\ set} \\
\\
10100 & \quad \left\{ \begin{array}{l}
\alpha g_2 '=c_1 +c_2 g_1 +c_4 g_2 \\
\alpha g_1 '=c_5 +c_6 g_1 \end{array} \right.
\\ & \\
11000 & \quad \left\{ \begin{array}{l}
\alpha g_2 '=c_1 +c_2 g_1 +c_3 \alpha g_1 '\\
g_2 =c_5 +c_6 g_1 +c_7 \alpha g_1 ' \end{array} \right.
\\ & \\
11100 & \quad \left\{ \begin{array}{l}
\alpha g_2 ' =c_1 +c_2 g_1 \\
g_2 =c_5 +c_6 g_1 \\
\alpha g_1 '=c_8 +c_9 g_1 \end{array} \right.
\end{array}
\nonumber
\label{smalltable}
\end{equation}

The goal of the algorithm is to find all sets of the free integrals
in order for the $\alpha\,,x$ variables to be separable.
If we know the free integrals as functions of $\alpha$, then it is
trivial to find the ODE's for the functions of $x$; we need to simply substitute 
$g_1\,,g_2$ in Eq.(\ref{singleeq}) and equate with zero the coefficients
of the independent functions of $\alpha$ in the resulting sum.
But before performing this step it is needed to clarify which are the possible 
forms of the free integrals; the solutions of Eq.(\ref{smalltable}).

For the family 10100:
For $c_6 \neq 0$ we have $g_1=-c_5/c_6+C \alpha^{c_6}$,
while for $c_6=0\,, g_1=c_5 \ln \alpha +C$.
In both cases we can find 
$g_2= \alpha^{c_4} \displaystyle \int \frac{c_1+c_2 g_1}{\alpha^{c_4+1}}d\alpha$.

Exploring this last equation we conclude that the case 10100 has
the following subcases (in the final expressions we may ignore additive constants
since the integrals depend only on the derivatives of $g_1\,,g_2$
while in the expression for the pressure, Eq.(\ref{pressuresum}),
the unknown $f_0$ may include the parts associated
with these constants):
\begin{eqnarray}
\mbox{1. If}  &
c_4 \neq 0\,, c_6 \neq 0 \,, c_4 \neq c_6\,,&
\left\{ \begin{array}{l}
g_1=-c_5/c_6+C \alpha^{c_6} \\
g_2=C \alpha^{c_4} + \frac{c_2}{c_6-c_4} \alpha^{c_6}
\end{array} \right. \nonumber \\
\mbox{2. If} &
c_4 \neq 0\,, c_6 \neq 0 \,, c_4 = c_6 \,,&
\left\{ \begin{array}{l}
g_1=-c_5/c_6+C \alpha^{c_6} \\
g_2=C \alpha^{c_4} + c_2 \alpha^{c_4} \ln \alpha
\end{array} \right. \nonumber \\
\mbox{3. If} &
c_4 = 0\,, c_6 \neq 0 \,,&
\left\{ \begin{array}{l}
g_1=-c_5/c_6+C \alpha^{c_6} \\
g_2=c_1 \ln \alpha + \frac{c_2}{c_6} \alpha^{c_6}
\end{array} \right. \nonumber \\
\mbox{4. If} &
c_4 \neq 0\,, c_6 = 0\,,&
\left\{ \begin{array}{l}
g_1=c_5 \ln \alpha +C \\
g_2=C \alpha^{c_4} - \frac{c_2 c_5}{c_4} \ln \alpha
\end{array} \right. \nonumber \\
\mbox{5. If} &
c_4 = 0\,, c_6 = 0 \,,&
\left\{ \begin{array}{l}
g_1=c_5 \ln \alpha +C \\
g_2=c_1 \ln \alpha +\frac{c_2 c_5}{2} (\ln \alpha)^2
\end{array} \right. \nonumber
\end{eqnarray}

Next we explore the 11000 and 11100 cases.
Each one has many subcases (as case 10100 has the above five).
We may simplify the final expressions noting that
\begin{itemize}
\item
Some subcase from one case may be the same with some other from another case.
\item
any additive constant in $g_1\,,g_2$ can be neglected; it is equivalent with
a rename of the unknown function $f_0$ in Eq.(\ref{singleeq}).
\item
Since $\alpha =G(x) \exp{(-Z/Z_0)}$ we may replace a power of $\alpha$ with $\alpha$;
this is equivalent with a rescaling of $Z_0$. So we are free to choose the exponent in a
possible term $\alpha^{\lambda}$. Of course if there are more than one terms of this
kind we can choose only one exponent.
\end{itemize}

Our final results for the free integrals and the corresponding ODE
with respect to $x$ are summarised in Table 1.
The expressions for the pressure in each case are also given.
\footnote{
There is another way to find all possible cases of Table 1.
Our goal is to find three ODE's (for the functions $f_0\,,M\,,G$) from Eq.(\ref{singleeq}).
So, all possible cases of the Table are answers to the question:
Find three independent vectors $\left[u_1(\alpha)\,,u_2(\alpha)\,,u_3(\alpha)\right]$
in the space of functions of $\alpha$, such that all the $Y_i$ of Eq.(\ref{Ys})
belong to the resulting subspace i.e. they are linear combinations of $u_1\,,u_2\,,u_3$.
We may choose $u_1(\alpha)=1$ (since unity is one of the $Y_i$'s)
and $u_2(\alpha)=g_1$ since $g_1$ cannot be a constant and thus it is independent
from $u_1$).}

\begin{table*}
\caption{Translationally self-similar exact MHD solutions}
$
\begin{array}{llll}
\quad&\quad {\rm \alpha-integrals} & \qquad \qquad \qquad {\rm ODE} & \qquad \qquad {\rm pressure} \\
\hline
1.\quad&
\begin{array}{l}
g_1=C_1 \alpha^2+C_2 \alpha^{\lambda} 
\\
g_2=D_1 \alpha^2+D_2 \alpha^{\lambda}
\end{array}
&
\begin{array}{l}
C_1 f_2 '-2 C_1 f_4 +D_1 f_3 ' - 2 D_1 f_5 =0 \\
C_2 f_2 '-\lambda C_2 f_4 +D_2 f_3 ' - \lambda D_2 f_5 =0 \\
f_0 '=0
\end{array}
&
\begin{array}{l}
\frac{4 \pi}{B_0^2}P=P_0+P_1 \alpha^2+ P_2 \alpha^{\lambda} \\
P_0=f_0 \\
P_1=C_1 f_2+D_1 f_3-2 C_1 f_1 \\
P_2=C_2 f_2+D_2 f_3-\lambda C_2 f_1
\end{array}\\
\hline
2.\quad&
\begin{array}{l}
g_1=C_1 \alpha^2+C_2 \ln{\alpha} \\
g_2=D_1 \alpha^2+D_2 \ln \alpha 
\end{array} &
\begin{array}{l}
C_1 f_2 '-2 C_1 f_4 +D_1 f_3 ' - 2 D_1 f_5 =0 \\
C_2 f_2 ' +D_2 f_3 '=0 \\
f_0 '-C_2 f_4 -D_2 f_5=0
\end{array} &
\begin{array}{l}
\frac{4 \pi}{B_0^2}P=P_0+P_1 \alpha^2+ P_2 \ln{\alpha} \\
P_0=f_0-C_2 f_1 \\
P_1=C_1 f_2+D_1 f_3-2 C_1 f_1 \\
P_2=C_2 f_2+D_2 f_3
\end{array} \\
\hline
3.\quad&
\begin{array}{l}
g_1=C_1 \ln \alpha+C_2 \left(\ln{\alpha}\right)^2 \\
g_2=D_1 \ln \alpha+D_2 \left(\ln{\alpha}\right)^2
\end{array} &
\begin{array}{l}
C_1 f_2 '-2 C_2 f_4 +D_1 f_3 ' - 2 D_2 f_5 =0 \\
C_2 f_2 ' +D_2 f_3 '=0 \\
f_0 '-C_1 f_4 -D_1 f_5=0
\end{array} &
\begin{array}{l}
\frac{4 \pi}{B_0^2}P=P_0+P_1 \ln \alpha +P_2 \left(\ln \alpha \right)^2 \\
P_0=f_0-C_1 f_1 \\
P_1=C_1 f_2+D_1 f_3-2 C_2 f_1 \\
P_2=C_2 f_2+D_2 f_3
\end{array} \\
\hline
4.\quad&
\begin{array}{l}
g_1=C_1 \alpha^2+C_2 \alpha^2 \ln{\alpha}\\
g_2=D_1 \alpha^2+D_2 \alpha^2 \ln \alpha
\end{array} &
\begin{array}{l}
C_1 f_2 '-\left(2 C_1+C_2\right) f_4 +D_1 f_3 ' -\left(2 D_1+D_2\right) f_5 =0 \\
C_2 f_2 ' -2 C_2 f_4 +D_2 f_3 ' - 2 D_2 f_5=0 \\
f_0 '=0
\end{array} &
\begin{array}{l}
\frac{4 \pi}{B_0^2}P=P_0+P_1 \alpha^2 + P_2 \alpha^2 \ln \alpha \\
P_0=f_0 \\
P_1=C_1 f_2+D_1 f_3-\left(2 C_1 +C_2\right)f_1 \\
P_2=C_2 f_2+D_2 f_3- 2 C_2 f_1
\end{array} \\
\hline
\end{array}
$
\label{largetable}
\end{table*}

\subsection{The subcases of some known solutions}

As mentioned in Section \ref{introduction} the systematic solution method described in the previous subsections is capable of recovering several well-known solutions as special cases of solutions of Table \ref{largetable} as well as new families first presented in this paper.  
In this section we give some details of how this work generalises certain specific solutions well-known in solar physics.

The first family of Table 1 is simplified if we choose
$C_2=D_2=f_0=0$. Then there is only one relation between the unknown functions
$G(x)\,, M(x)$ and we may assume a polytropic relationship between pressure and density.  
Thus, by imposing as a second relationship $ P_1 = D_0 M^{-2 \gamma}$, and since 
$1/M^2 = 4\pi \rho /\Psi_A^2$, $\Psi_A^2 \propto \alpha^2$, it follows 
that $P = \frac{B_0^2}{4 \pi}D_0 \alpha^2 M^{-2 \gamma} = Q(A) \rho^{\gamma}$.
The isothermal subcase ($\gamma=1$) has been analyzed in
Tsinganos, Surlantzis and Priest (1993).
Besides that special subcase, we recover the general non-isothermal 
polytropic case as a subcase of the first family in Table 1.

The ``given temperature profile'' solutions of Hood \& Anzer (1990) belong to
the first family of Table 1 also, with $C_2=D_2=f_0=0$. The second relation between 
$G(x)\,, M(x)$ is the given temperature profile (the product $P_1 M^2$, which is
proportional to the temperature, is a given function of $x$).  This is also the case with Del Zanna \& Hood (1996), whose generalisation of Hood \& Anzer's model to include compressible flow is similar in method to Tsinganos, Surlantzis and Priest (1993) and is an identical special case of our scheme with this different imposed closure condition.

The classical Kippenhahn \& Schl\"uter (1957), ``prominence''-like solution belongs to the second family of Table 1,
assuming further $C_1=D_1=P_2=0$. If this is the case, $A \propto \ln \alpha\,,
\Psi_A = const$, while if this constant is zero we find static solutions.
Again there is only one relation between the unknown functions
$G(x)\,, M(x)$ and we can add a polytropic relationship between pressure and density
as the second one.

\section{A new class of solutions}
\label{newclass}

\begin{figure}
\includegraphics[width=0.5\textwidth]{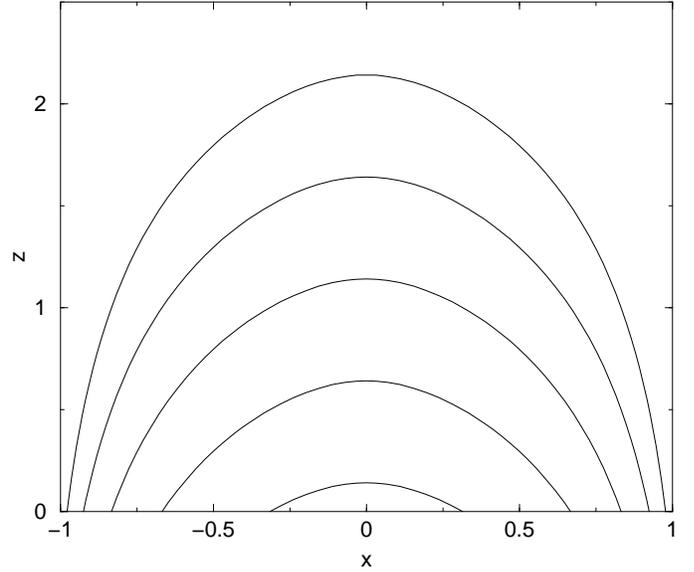}
\caption{Sample magnetic and flow field lines for the sub-Alfv\'enic example.  All field lines are of a loop-like shape and, because of the self-similarity of the solution, the lines differ from each other only by an additive constant.  This solution has physical limits on the $X$-axis defined by a limit imposed on the plasma pressure and density (plotted in Fig. \ref{subparams}) so that they do not become unphysically large. The horizontal and vertical coordinates are given in 
units of $Z_0$, $x=X/Z_0$ and $z=Z/Z_0$.}
\label{subfieldlines}
\end{figure}

\begin{figure}
\centering
\includegraphics[width=0.5\textwidth]{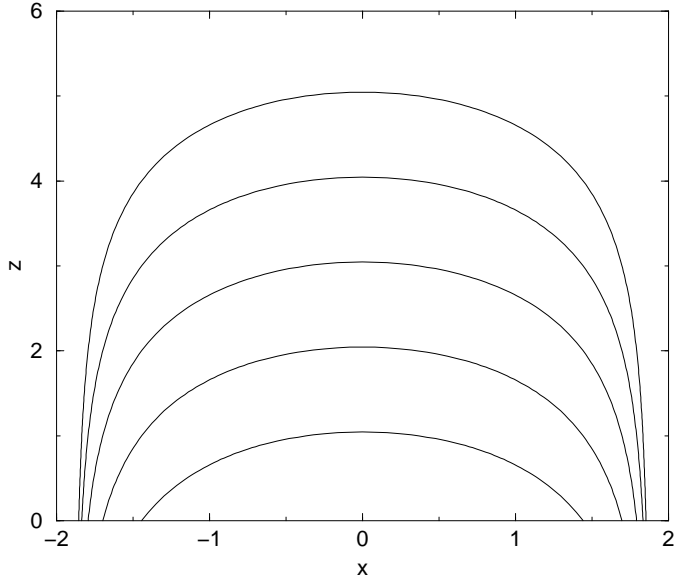}
\caption{Sample magnetic and flow field lines for the trans-Alfv\'enic example.  All field lines are of a loop-like shape and, because of the self-similarity of the solution, the lines differ from each other only by an additive constant.  This solution has physical limits on the $X$-axis defined by the condition that the derivative $F'(x)$ of the slope of a line must be negative.  At the foot points of the largest loop this derivative is zero and the line is exactly straight but reverse curvature is avoided. The horizontal and vertical coordinates are given in 
units of $Z_0$, $x=X/Z_0$ and $z=Z/Z_0$.}
\label{transfieldlines}
\end{figure}

\begin{figure*}
\centering
\includegraphics[width=0.49\textwidth]{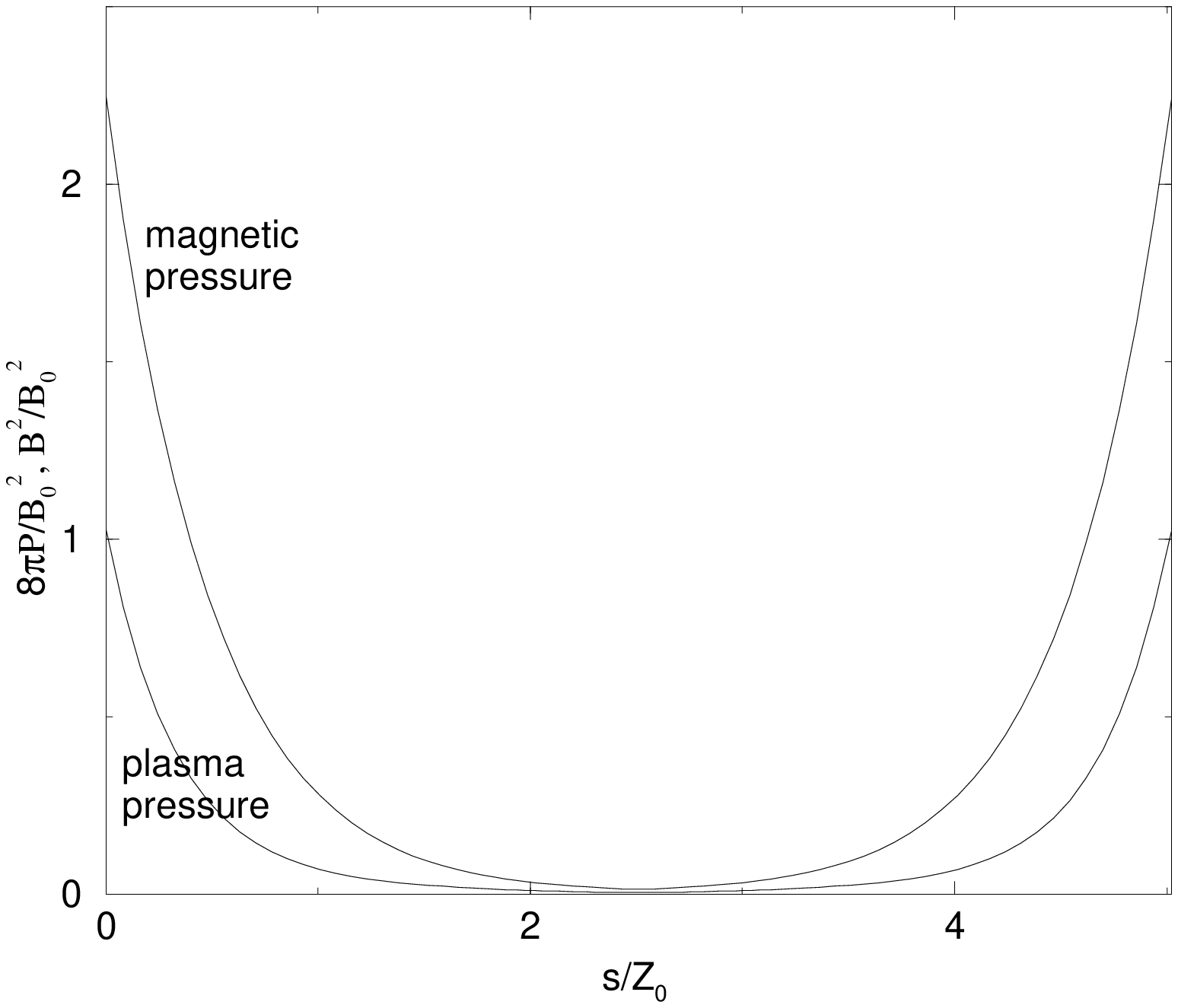}
\includegraphics[width=0.49\textwidth]{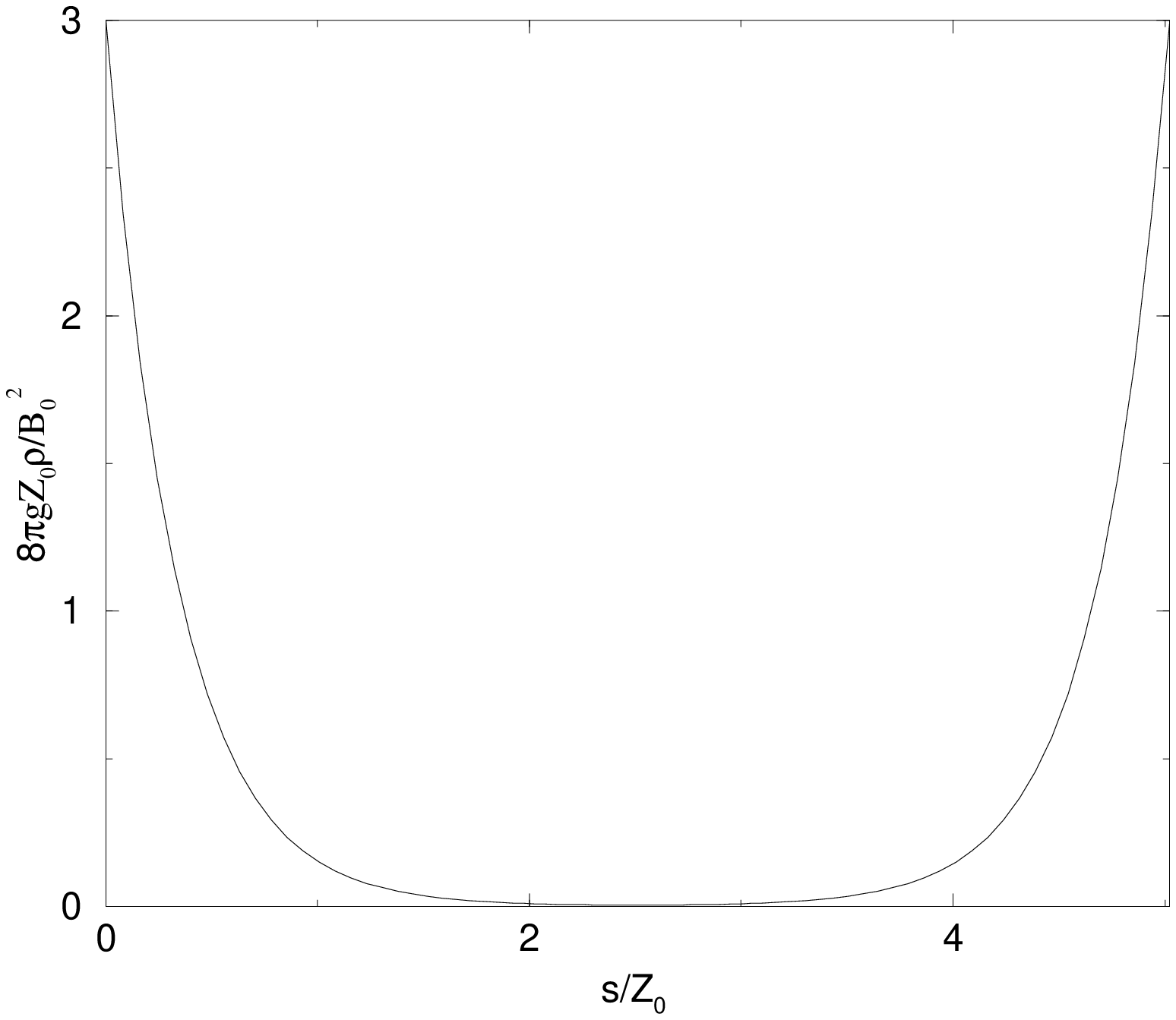}
\includegraphics[width=0.49\textwidth]{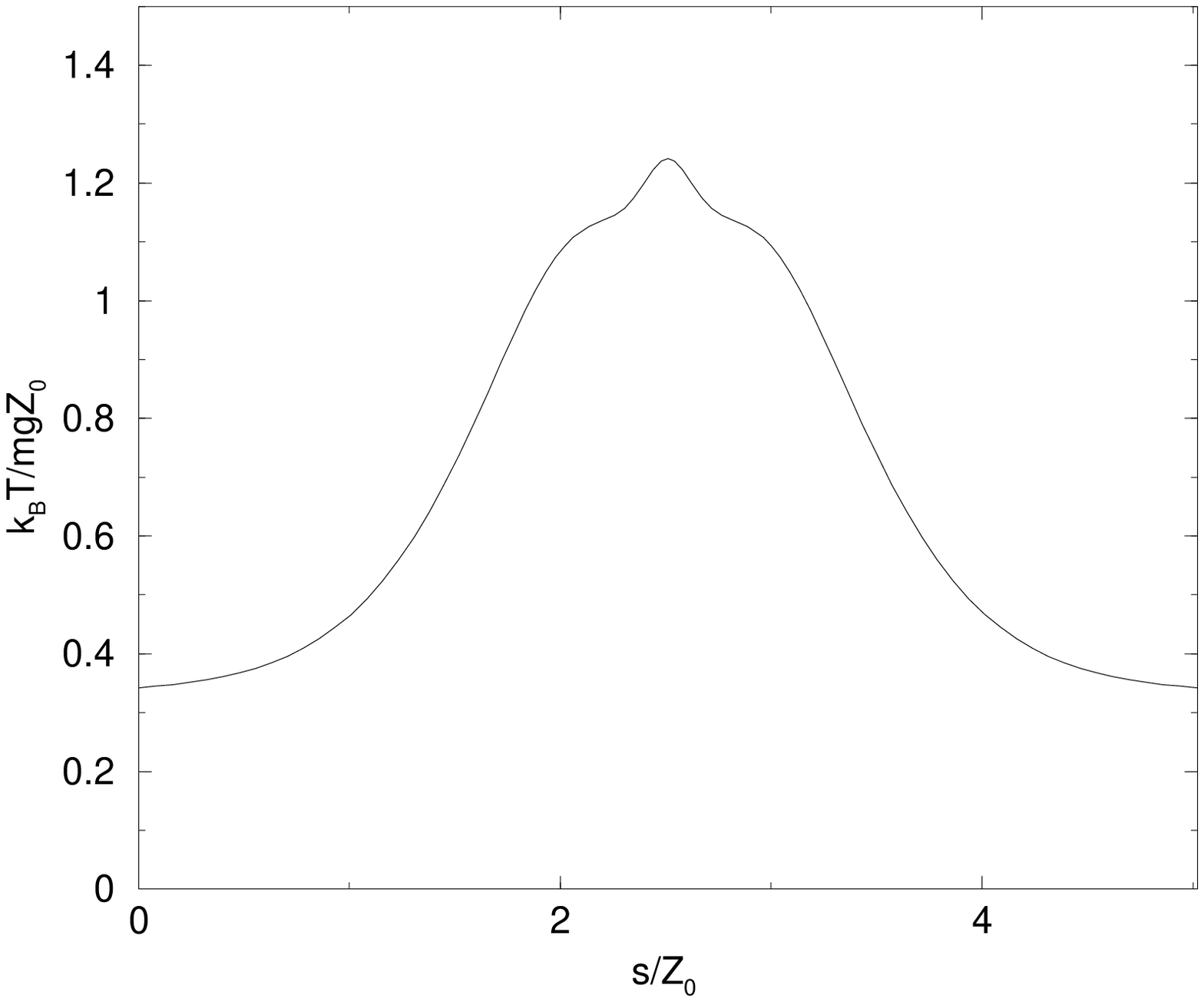}
\includegraphics[width=0.49\textwidth]{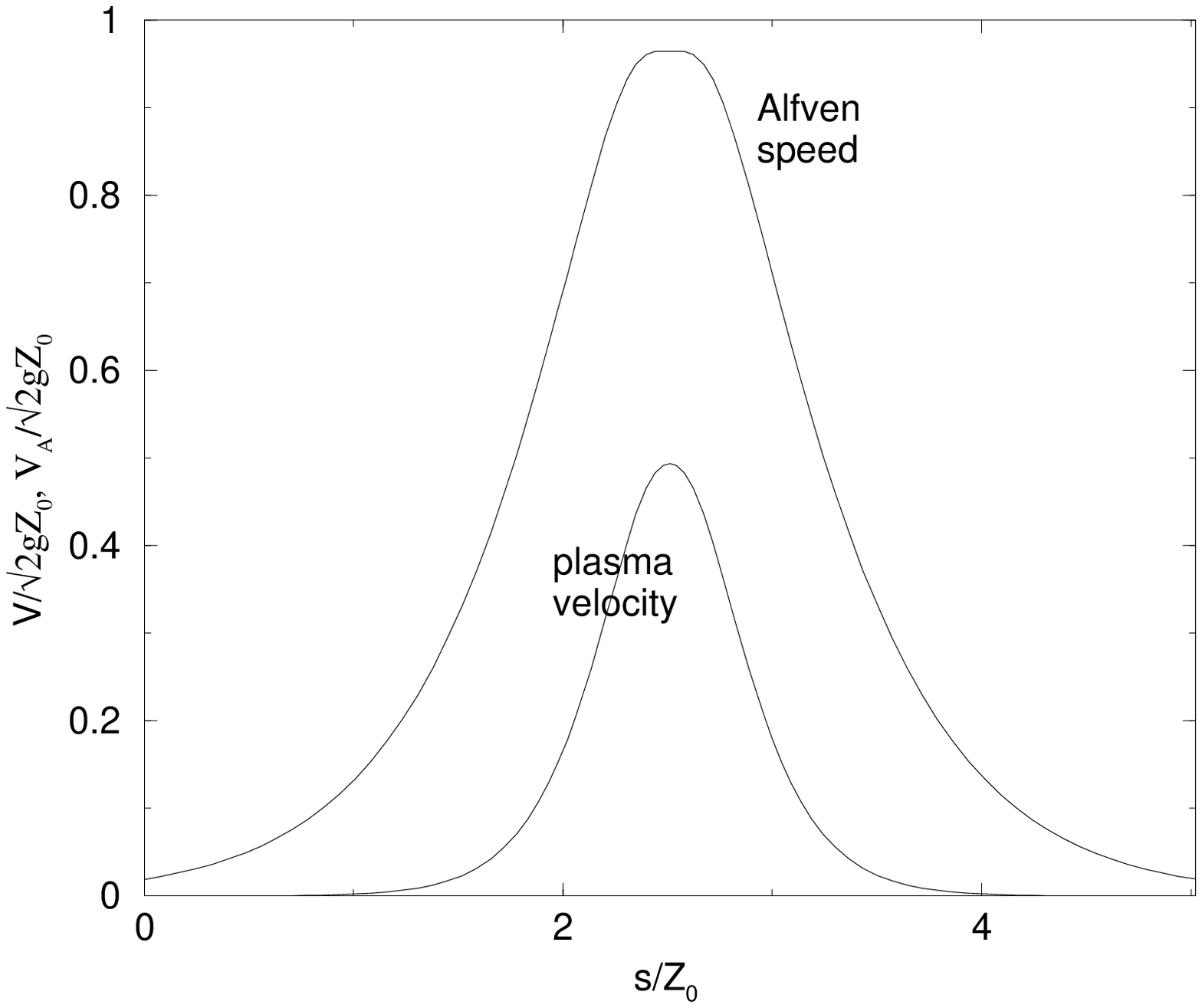}
\caption{Graphs of plasma quantities for the sub-Alfv\'enic example: plasma and magnetic pressures (top left), plasma density (top right), plasma temperature (bottom left) and plasma velocity and Alfv\'en speed (bottom right) along the largest loop shown in Fig. \ref{subfieldlines}.  All quantities are graphed against the arc length $s$ of this loop, taking as $s=0$ the left foot point.  The U-shape profiles for the density and pressure reflects the fact that the density and pressure decrease with height, while the temperature and velocity increase with height along the loop.  Note that the velocity is clearly sub-Alfv\'enic and that the magnetic pressure clearly exceeds the plasma pressure.}
\label{subparams}
\end{figure*}

It seems that a particularly interesting new case is the first family in Table 1.
The non-polytropic subcase,
$|C_2|+|D_2| \neq 0$,
has two ``scales'': $Z_0$ and $2 Z_0/\lambda$.
In this section we examine this case, with $C_2=0$.

The corresponding integrals are
\begin{eqnarray}
&& g_1=C_1 \alpha^2 \,, \ \ g_2 = D_1 \alpha^2 + D_2 \alpha^{\lambda} \,, \quad \\\mbox{or,} 
\nonumber \\
&& A=Z_0 B_0 \sqrt{2 C_1}\ \alpha \,, \Psi_A^2=\frac{B_0^2}{g Z_0}
\left(2 D_1 \alpha^2 + \lambda D_2 \alpha^{\lambda}\right) \,.
\end{eqnarray}
Note that we may choose $C_1$ such that $B_0$ is a component of the magnetic field at a 
reference point. The form of the pressure is 
\begin{eqnarray}
P =\frac{B_0^2}{4 \pi} \left(P_0+P_1 \alpha^2+ P_2 \alpha^{\lambda}\right) \,,
\nonumber
\end{eqnarray}
where
\begin{eqnarray}
P_1&=&C_1 \left[FM^{2'}- F^{'} (1-M^2)-F^2-1\right]+\frac{D_1}{M^2} \,,
\nonumber \\
P_2&=&\frac{D_2}{M^2} \,, \quad P_0=f_0\,.
\nonumber
\end{eqnarray}
Using the above definitions for the ``pressure components'' together with the
ODE's from Table 1, we conclude that for the nontrivial case
$\lambda \neq 0 \,, \lambda \neq 2\,, D_2 \neq 0\,,$
we have the following system of equations for the six unknown functions of $x$
(including the slope of the field lines $F$):

\begin{eqnarray}
P_0 & = & {const}\,, \label{P0}\\
P_2 & = & \frac{D_2}{M^2}\,, \label{P2}\\
\frac{d \ln G}{dx} & = & F\,, \label{F}\\
\frac{d \ln M^2}{dx}& = & \lambda F\,, \ \mbox{or,} \  M^2 = D_3 G^{\lambda}\,, \mbox{with} \ D_3=const\,, \label{lambdaF}\\
F '& = & \frac{\left(\lambda M^2 -1\right) F^2 -1 + \displaystyle \frac{D_1}{C_1M^2}-
\displaystyle \frac{P_1}{C_1}}
{1-M^2} \,, \label{Fd}\\
P_1 ' & = & -\frac{2 D_1 F}{M^2} - 2 C_1 F M^2 \left[
F '+\lambda \left(F^2+1\right)\right]\,.\label{P1d}
\end{eqnarray}

Note that practically the solution for the shape of the loop is a three-parametric 
family depending on the constants $D_1/C_1$, $\lambda$ and $D_3$.
These ODE's can be solved by standard numerical methods.  Eqs. (\ref{F},\ref{Fd},\ref{P1d}) form a closed system of ODE's in the three variables $G$, $F$ and $P_1$.
On solving these, $M^2$ and $P_2$ can be recovered from the remaining equations.
We remark that in the case $D_1/C_1\neq 0$, $F'$ and $P_1'$ become large for small $M$ and the solution becomes unphysical.
In order to obtain a loop-like solution it is necessary to keep $F'$ negative.
If the solution is super-Alfv\'enic near the apex of the loop where $F\approx 0$ then $F'$ can only be kept negative by having $D_1/C_1\neq 0$.
A strictly sub-Alfv\'enic solution can, however, have a loop-like appearance with
$D_1/C_1=0$ as will be seen, although we cannot save the pressure and density
from becoming infinite as $x\rightarrow\pm\infty$.
From Eq.(\ref{lambdaF}) with $\lambda >0$ the Alfv\'en Mach number graph resembles a typical
loop-like field line,
an inverted U symmetric about $x=0$, but always strictly positive.
From Eq.(\ref{M2}) $\rho\propto {1 / M^2}$ along a field line and therefore $\rho$ tends to $\infty$ as $x\rightarrow\pm\infty$, as does $P_2$ from Eq.(\ref{P2}).
In the sub-Alfv\'enic case it is possible to control $P_2$'s contribution to the solution by sending $D_2\rightarrow 0$
but there is little advantage to be gained from doing this.  $P_1$ also becomes infinite because of the dependence of $P_1'$ on $F$, which becomes infinite if $F'$ is kept negative.

We would like the Alfv\'en speed $V_A=|{\bf B}|/\sqrt{4\pi\rho}$ to increase with height and the plasma $\beta =8\pi P/|{\bf B}|^2$ to decrease with height, as is the case in the solar atmosphere, and this can be guaranteed by setting $\lambda >2$.  We remark that an alternative approach is to set $0<\lambda <2$ and $D_2<0$.  We refrain from doing this since using a negative pressure ``component'' introduces complications in ensuring that the pressure is everywhere non-negative and decreasing with height.
\footnote{the above requirements are equivalent to $\lambda>0$ and $(\lambda-2)D_2>0$. So,
there are two possible cases: $\lambda>2\,, D_2>0$ or $0<\lambda<2\,, D_2<0$.}

\begin{figure*}
\centering
\includegraphics[width=0.49\textwidth]{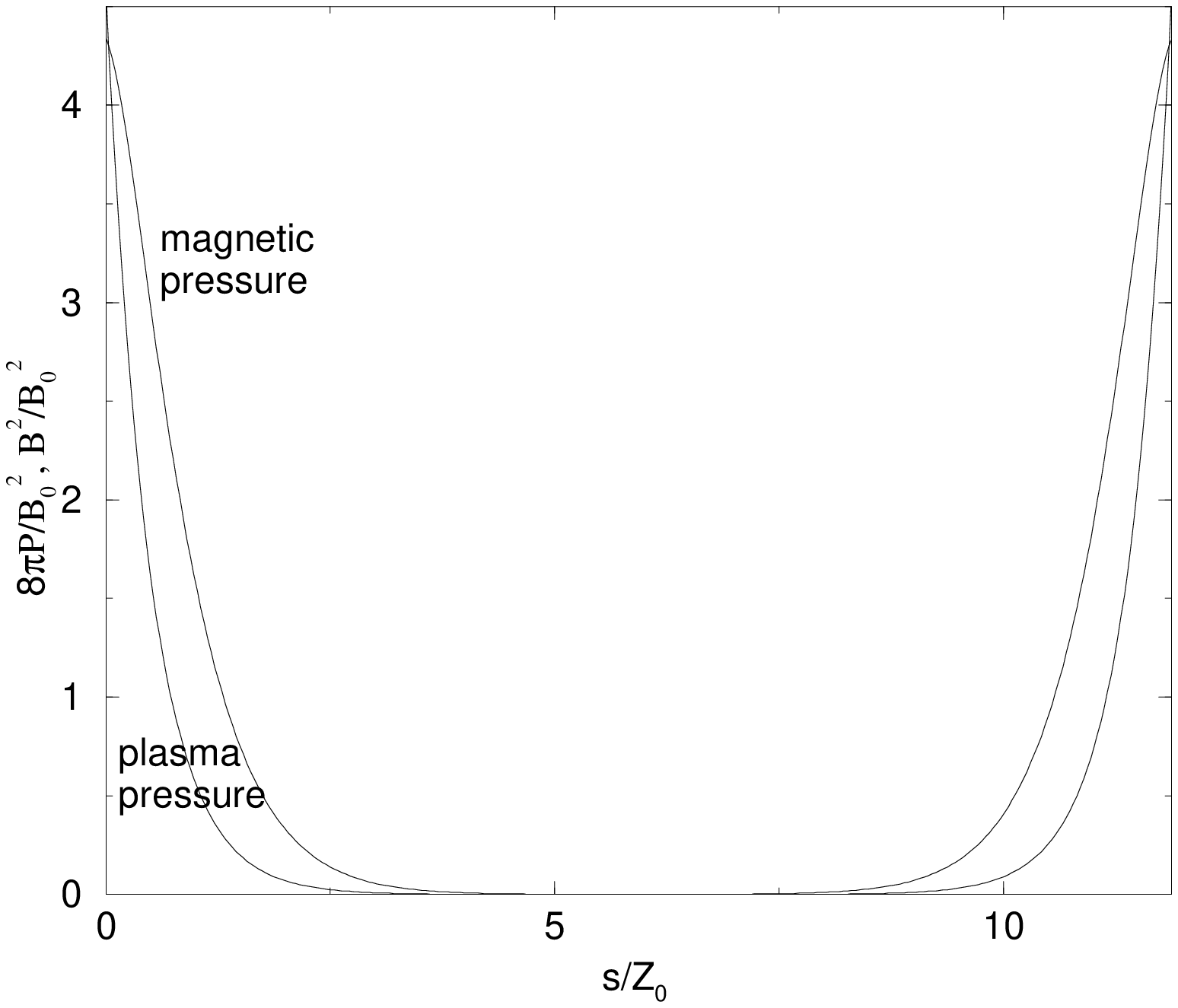}
\includegraphics[width=0.49\textwidth]{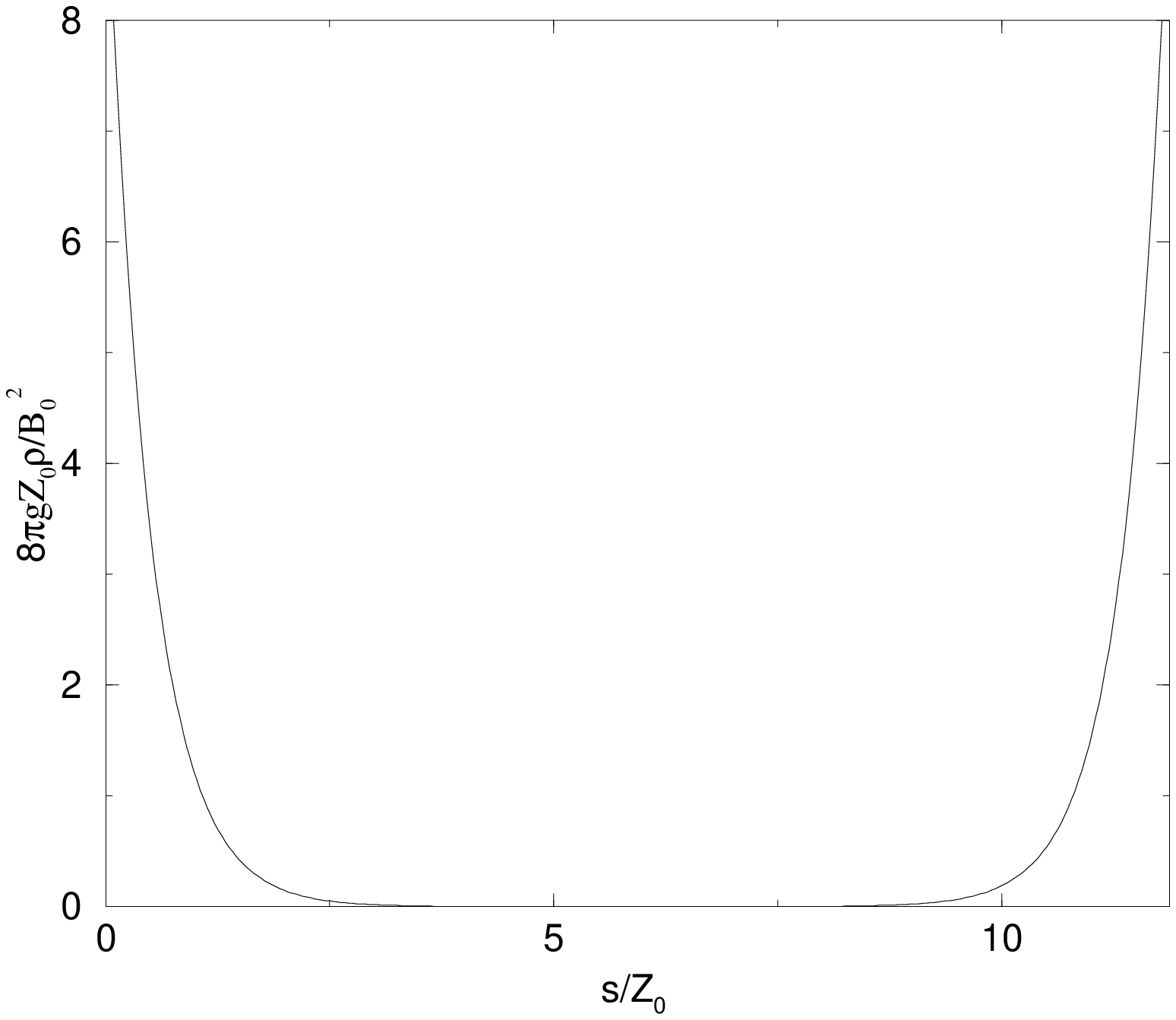}
\includegraphics[width=0.49\textwidth]{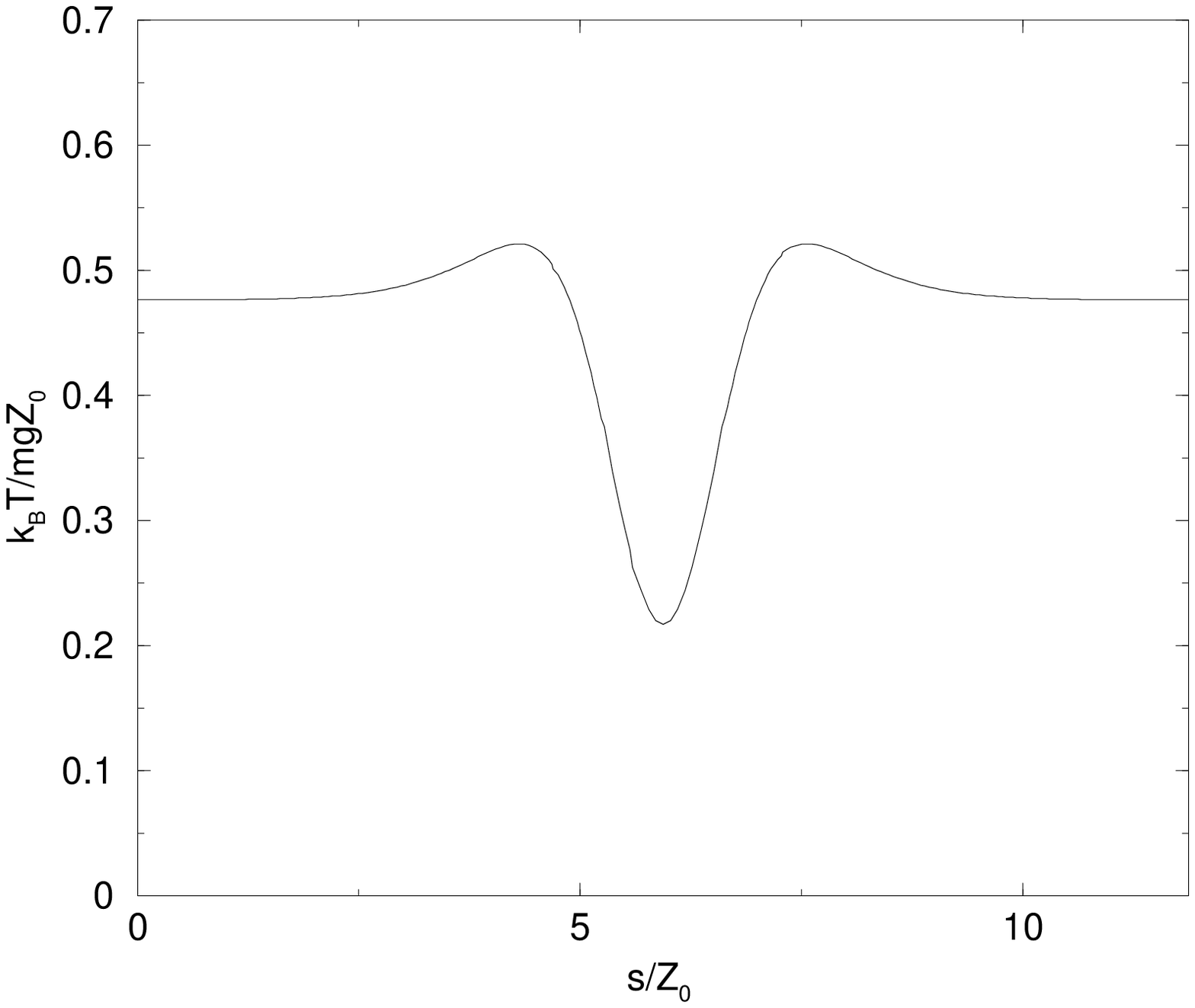}
\includegraphics[width=0.49\textwidth]{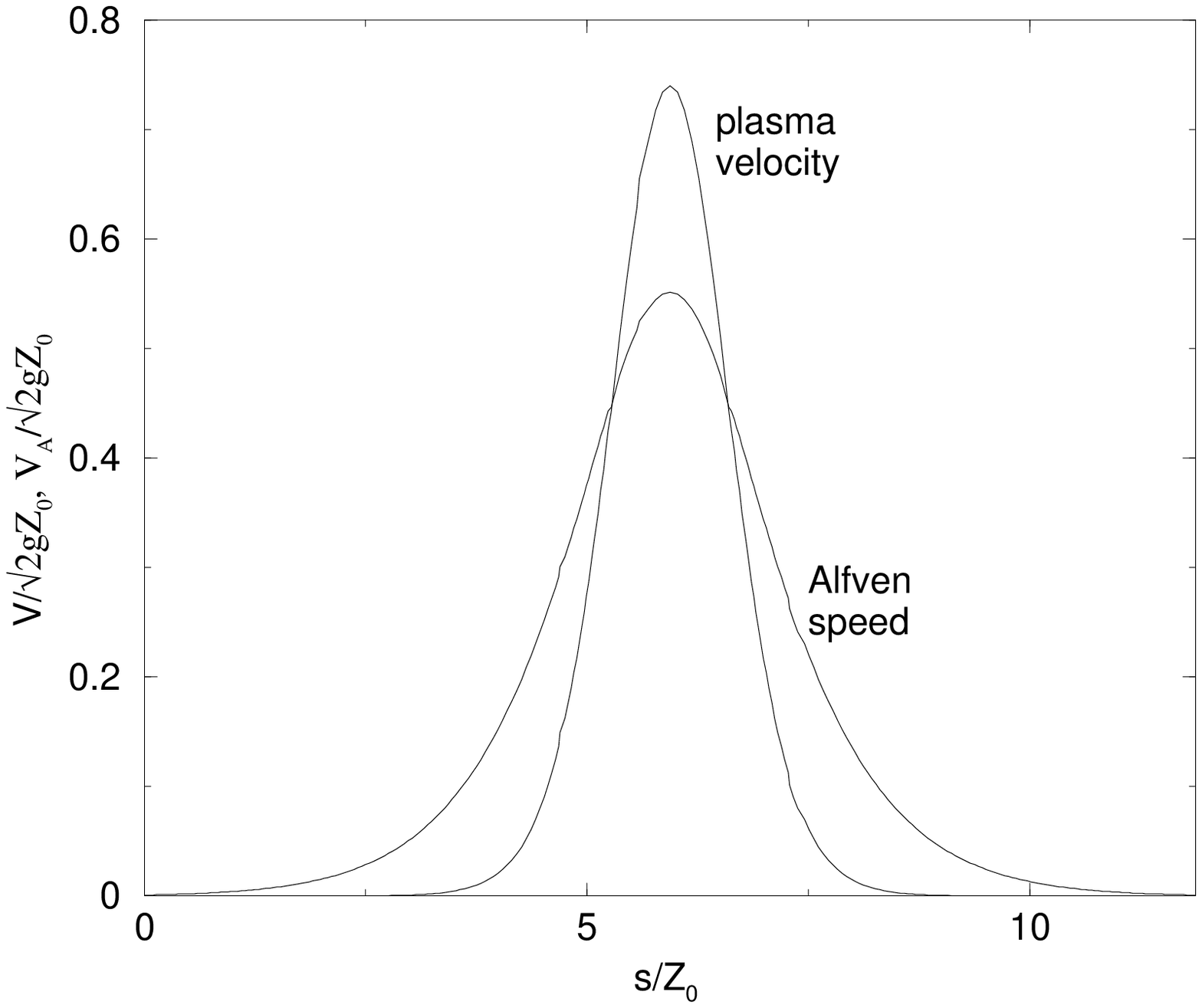}
\caption{Graphs of plasma quantities for the trans-Alfv\'enic example: plasma and magnetic pressures (top left), plasma density (top right), plasma temperature (bottom left) and plasma velocity and Alfv\'en speed (bottom right) along the largest loop shown in Fig. \ref{transfieldlines}.  All quantities are graphed against the arc length $s$ of this loop, taking as $s=0$ the left foot point.  The U-shape profiles for the density and pressure reflects the fact that the density and pressure decrease with height, while the velocity increases with height along the loop.  Note that the magnetic pressure is greater than the plasma pressure, and the velocity is clearly trans-Alfv\'enic, being super-Alfv\'enic near the apex and sub-Alfv\'enic near the foot points, and that the critical points are crossed smoothly.  In contrast to the sub-Alfv\'enic example of Fig. \ref{subparams} the temperature decreases with height along the super-Alfv\'enic part of the loop.}
\label{transparams}
\end{figure*}

We generate loop-like solutions as follows.  We begin by calculating the right half of the loop, beginning from the loop apex at $x=0$ with suitable initial conditions $F=0$, $G>0$ and $P_1>0$.  For simplicity we assume throughout that $C_1=1.0$.
The symmetry properties of Eqs. (\ref{P0} - \ref{P1d}) ensure that on integrating from $x=0$ in the negative direction the other half of a symmetric loop-like solution is obtained.
In the sub-Alfv\'enic case the equations have no critical points and can be integrated without difficulty.
Figure \ref{subfieldlines} shows some example field lines of our sub-Alfv\'enic example and the Alfv\'en Mach number profile.  The parameter values used in this example are $D_1/C_1=0.0$, $D_2=1.0$,
$D_3=1.0$ and $\lambda =3.0$.  Figure \ref{subparams} shows graphs of the plasma pressure and magnetic pressure (top left), the plasma density (top right), the plasma temperature (bottom left) and the plasma velocity and Alfv\'en speed (bottom right).

In the trans-Alfv\'enic case, however, Eq.(\ref{Fd}) and therefore Eq.(\ref{P1d}) have critical points where $M^2=1$.  
In order to obtain a regular trans-Alfv\'enic solution it is necessary to fix the parameters so that Eq.(\ref{Fd}) has no singularity at $M^2=1$.  The relation

\begin{equation}
(\lambda -1)F_{\star}^2-1+{D_1\over C_1}-{P_{1\star}\over C_1}=0 \label{regularity}
\end{equation}

\noindent 
must hold, where a star denotes evaluation of a quantity at the Alfv\'en point $M^3=1$.
Since Eq.(\ref{regularity}) involves $F_{\star}$ and $P_{1\star}$ the parameters $D_1/C_1$, $D_3$ and $\lambda$ cannot be fixed to solve Eq.(\ref{regularity}) independently of the solution.
The correct combination of parameter values can be found by a shooting method as seen in Vlahakis \& Tsinganos (1999).  
We begin the integration with a reasonable set of parameter values and, as the solution approaches the Alfv\'en point the solution will asymptote and will approach $+\infty$ or $-\infty$  depending on the sign of the numerator of 
Eq.(\ref{Fd})'s right-hand side at this point.
Bearing in mind that a regular solution will have numerator zero at the Alfv\'en point we adjust a chosen 
parameter iteratively until we have a regular solution.
It is possible to produce a regular solution avoiding this iterative procedure by starting the integration  
at the Alfv\'en point and imposing finite values there, but it is difficult to impose symmetry on the loop 
using this approach.  We find it more convenient to begin the integration at the apex, imposing symmetry 
via the initial value of the slope, and iterate to find a solution regular at the Alfv\'en point.  
The parameter values used in the trans-Alfv\'enic example given here are $D_1/C_1=2.20659184272718$, $D_2=1.0$,
$D_3=1.0$ and $\lambda=2.1$.

Example field lines for the two examples are shown in Figs. \ref{subfieldlines} and \ref{transfieldlines}.  
The field lines are all loop-like in shape and the translational self-similarity of the solutions causes any two 
field lines in a solution to differ from each other only by a constant value in the $z$-direction.  The slightly 
different shape of field line between the two examples is due to the different values of $D_1/C_1$ and $\lambda$.  
In Figs. \ref{subparams} and \ref{transparams}, all quantities can be seen to have symmetric profiles.  
The pressure, density and magnetic field strength along the field line all have their minima at the apex and 
increase monotonically towards the foot points since these quantities all decrease exponentially with height.  
Note that because $\lambda >2$ and $D_2>0$ the velocity in this example is an increasing function of $z$, 
as is the Alfv\'en speed (see Figs. \ref{subparams} and \ref{transparams}, bottom right pictures), 
and the plasma $\beta$ is a decreasing function of $z$,  as well as all being dependent on $x$.  
This is all in contrast to Tsinganos, Surlantzis \& Priest (1993)'s example where the velocity, Alfv\'en speed 
and plasma beta are all functions of $x$ alone.  (Of course the Alfv\'en Mach number depends on $x$ alone 
both here and in Tsinganos, Surlantzis \& Priest (1993).)

\begin{figure*}
\centering
\includegraphics[width=0.49\textwidth]{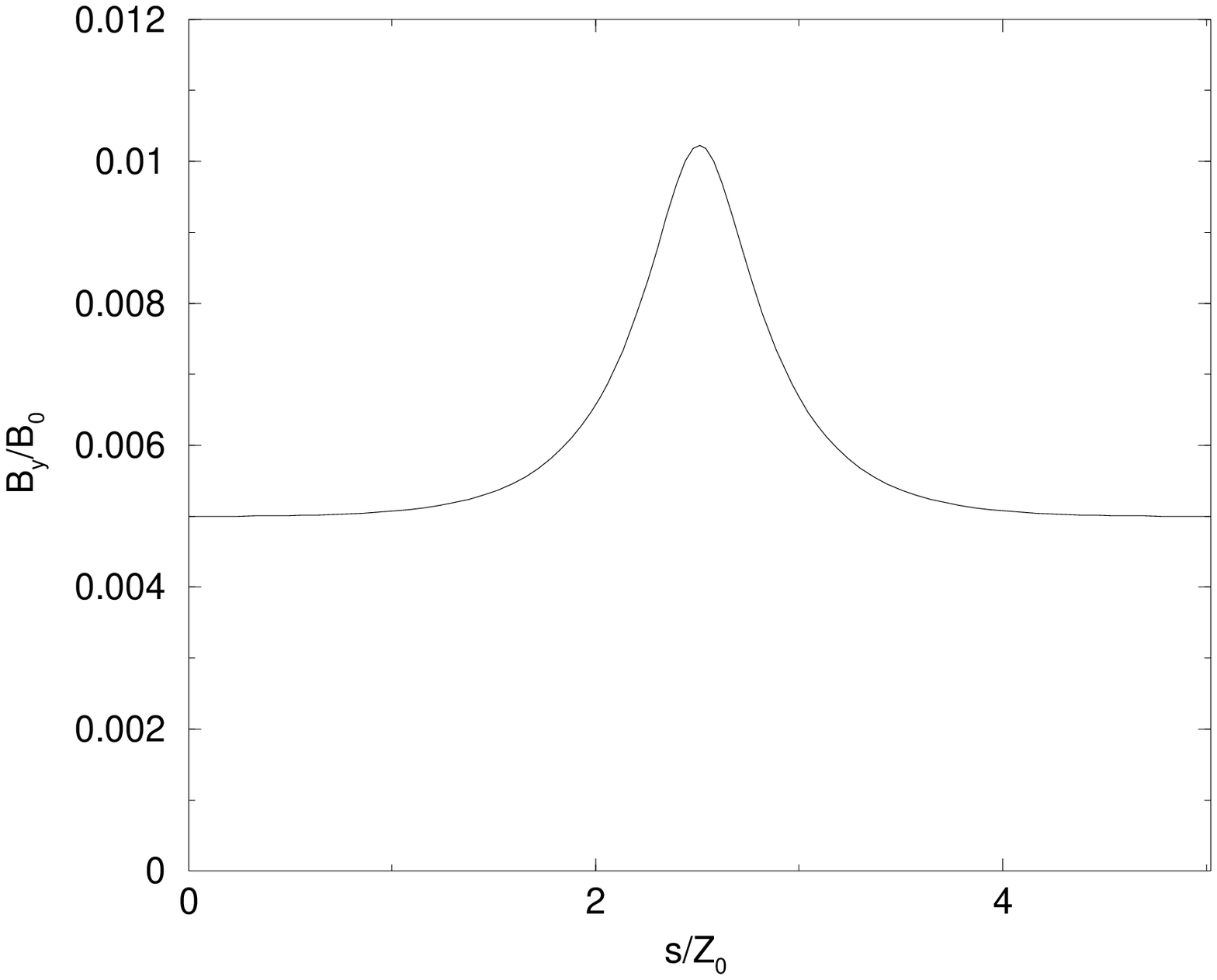}
\includegraphics[width=0.49\textwidth]{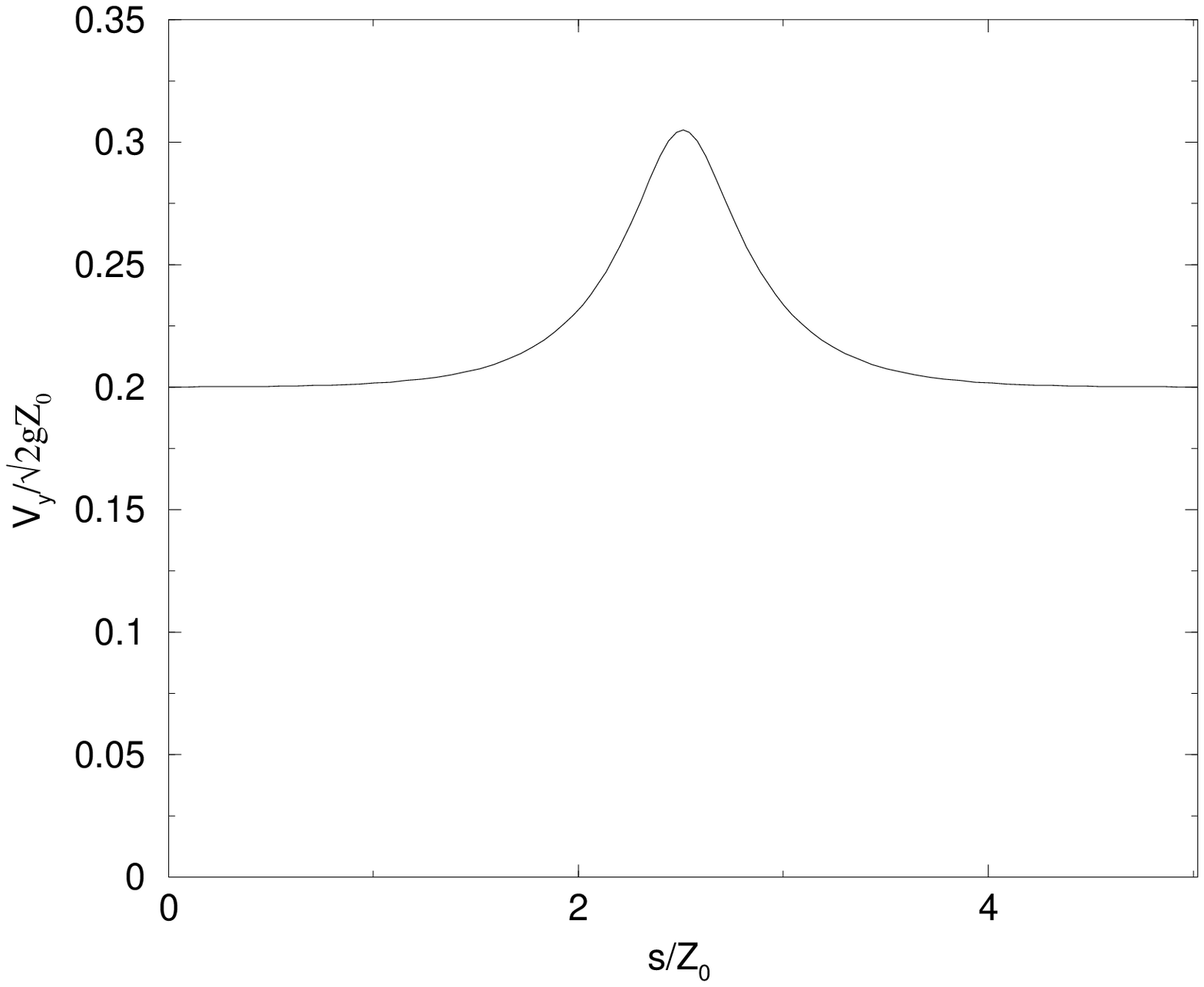}
\caption{Plots of the y-components of the magnetic field (left picture) and the velocity field (right picture) for the sub-Alfv\'enic example graphed against the arclength of the largest field line of Fig. \ref{subfieldlines}.  In this example the free integrals are given the values $\Phi_A (A)=0.2$ and $L(A)=0.1$.  For both fields the shear is concentrated near the apex although, since $B_y$ and $V_y$ are not directly proportional, the flow is not field-aligned in this (or any non-planar) case.}
\label{ByVy}
\end{figure*}

Recall from Subsection \ref{assumptions} that non-planar components of the solution are also possible.  In the trans-Alfv\'enic example these components are simple: $B_y$ must be zero while $V_y$ is a free function of $\alpha $ and so must be constant on a particular field line but can vary arbitrarily from field line to field line.  In our discussion of trans-Alfv\'enic solutions we confine our attention to the planar case $B_y=V_y=0$ so that definitions of Alfv\'en points $M=1$ refer to velocities confined to the $x$-$z$ plane.  In the sub-Alfv\'enic case the absence of critical points permits more complicated non-planar magnetic fields and velocity fields.  The non-planar components $B_y$ and $V_y$ must satisfy Eqs. (\ref{ycpts}), where $\Phi_A(A)$ and $L(A)$ are free functions of $\alpha $ which are therefore constant on a particular field line but can vary arbitrarily from field line to field line.  Examples of possible profiles of $B_y$ and $V_y$ for our sub-Alfv\'enic example of Figs. \ref{subfieldlines} and \ref{subparams} are given in Fig. \ref{ByVy}.  In these plots $B_y$ (left picture) and $V_y$ (right picture) are graphed against the arclength of the largest field line of Fig. \ref{subfieldlines} and the values assigned to the free integrals are $\Phi_A (A)=0.2$ and $L(A)=0.1$.  With these values for the free integrals the shears of both the magnetic field and the velocity field are concentrated near the apex, but $B_y$ and $V_y$ are not directly proportional and so the flow is not field-aligned.  Note that in any case with or without Alfv\'en points the only possibility for field-aligned flow is the planar case $L=\Phi_A=0$.

Apart from the existence of critical points in this trans-Alfv\'enic example the examples integrate slightly differently from each other.  In the sub-Alfv\'enic case $G$ begins from its maximum at the apex of the loop and then tends monotonically to $0$.  In the trans-Alfv\'enic case, however, the non-zero value of $D_1/C_1$ and the decreasing value of $G$ (and hence of $M^2$) causes the term $D_1/C_1M^2$ in the numerator of Eq.(\ref{Fd}) to become large, $F'$ to become positive
and the loop to loose its loop-like shape.  We must confine our attention to the region where $F'$ is negative if we want a loop-like solution.  Adjusting $\lambda$ (and therefore $D_1/C_1$) varies the location of this point $F'=0$ and its value of $G(x)$.  This difference in behavior between the two examples affects our method of choice of $\lambda$.

In this Euler-potential formulation a field line can be defined as a line of constant $\alpha $.  We show graphs of quantities against loop arc length along a single field line for each solution and we choose our representative field line as follows.  In the sub-Alfv\'enic example where $D_1/C_1=0$ we choose an interval of the $x$-axis $[-x_0,x_0]$ over which the plasma quantities are of a reasonable size, i.e. where $M^2$ is not close enough to zero for $P$ or $\rho$ to be unphysically large.  In the trans-Alfv\'enic example where $D_1/C_1\neq 0$ we take the interval $[-x_0,x_0]$ to be the domain where $F'$ is strictly negative.  We take as our representative field line the line with foot points at the ends of this interval $x=\pm x_0$.  This line is defined by the equation $\alpha =G(x_0)=G_0$ or equivalently $z=\ln G(x) -\ln G_0$ (note that the translational self-similarity of the solution is explicit in this equation for the field lines).  
The patterns vary slightly from field line to field line but examination of several field lines in each example 
indicates that our chosen field line is broadly representative in each case, apart from the decrease of the plasma $\beta$ with increasing height.  In the trans-Alfv\'enic example this variation of $\beta$ limits our possible choice of values for $\lambda$.  Because $\beta$ decreases as $\alpha$ increases, if our solution is to include a low-$\beta$ field line it must extend far enough along the $x$-axis for small values of $G(x)$ (and therefore $\alpha $) to be included.  We find that for the present class of solutions $\lambda$ must be taken close to $2.0$ to find field lines with low $\alpha$ and therefore low $\beta$ values.  This is in sharp contrast to the sub-Alfv\'enic case where $F'\neq 0$ away from the apex and a wide range of low-$\beta$ field lines is available for any value of $\lambda$.

Hence in the two examples given here the plasma pressure is smaller than the magnetic pressure (see Figs. \ref{subparams} and \ref{transparams}, top left pictures).  Given that this is true in particular at the apex of the loop where $F=0$ and the value of $G$ is determined by our wanting a sub-Alfv\'enic solution or a trans-Alfv\'enic solution, control $\beta$ here by adjusting the value of $P_1$ at the apex as well as adjusting $P_2$ here via $D_2$.  To keep things simple in this initial study we prefer to keep the values of the variables at the apex and the values of the constants consistent between the two solutions apart from altering $G$ at the apex to obtain one sub-Alfv\'enic and one trans-Alfv\'enic example, using the low value of $\lambda$ in the trans-Alfv\'enic and adjusting $D_1/C_1$ to make the trans-Alfv\'enic solution regular.  Various start values of the variables and various values of the constants were tried and the results given here are 
as physically acceptable as any found.

A further difference between the two solutions is that their temperature profiles are very dissimilar (see Figs. \ref{subparams} and \ref{transparams}, bottom left pictures).  Because $\alpha $ is constant along a field line and because of Eq.(\ref{P2}) the expression for $T$ along a field line is of the form $K_1P_1(x)M^2(x)+K_2$
where $K_1$ and $K_2$ are constant.  The slightly different Alfv\'en Mach number profiles for the two examples (reflected in the slightly different field line shapes) and the slightly different profiles of the pressure ``component'' $P_1(x)$ have resulted in very different temperature functions for the two examples.  In the sub-Alfv\'enic example there is a clear temperature maximum at the apex of the loop and minima at the foot points, while in the trans-Alfv\'enic example the apex is the location of the temperature minimum and the temperature maxima are close to the foot points.  From the investigation of many examples it seems that, broadly speaking for this special subclass of solutions, temperature maxima are characteristic of sub-Alfv\'enic apexes and temperature minima are characteristic of super-Alfv\'enic apexes.  Near an apex where $M^2>1$ and $F$ is small, the $\lambda$ term in Eq.(\ref{P1d}) dominates so that $T$'s slope has the same sign as $P_1'$ causing $T$ to have a minimum there.  However, choosing a large value for the constant pressure ``component'' $P_0$, set to $0$ in these examples, affects the temperature profile and can even change local minima on the apex to maxima on field lines of small $\alpha $, although a large value for $P_0$ affects the plasma $\beta$ especially at large heights.  This discussion applies only to the special case $C_2=0$ of our first family of solutions.
A more general study of the relationship between field line shape and the thermodynamics of the more general case $C_2\neq 0$ and of other solution families will be the subject of a further paper but these two simple examples hint at how much variety is possible.

\section{Conclusions}
\label{conclusions}

We have presented a systematic method for constructing exact two-dimensional MHD equilibria with compressible flow in Cartesian geometry with a view to modelling structures in the solar atmosphere.  Using this method we have found several new classes of solutions, some of which include some well-known solutions as special cases and some of which are completely new.  As examples illustrating the method we have given two simple loop-like solutions, one with strictly sub-Alfv\'enic flow and one with trans-Alfv\'enic flow.  The critical points of the trans-Alfv\'enic example are crossed smoothly and all quantities in the model are regular.  The models have physical properties not previously available in exact two-dimensional MHD work.  Meanwhile two classes of solutions generalise previously-known loop and prominence models.  It is shown how an analytical description of heat input/output may be obtained for a solution whether polytropic or non-polytropic.  
It therefore seems likely that the solution construction method may be applied in future in providing quantitative 
models of prominences and arcades to be compared with observations or giving details of the heating profile of the structure.  Such applications will be the subject of future work.

\begin{acknowledgements}
GP acknowledges funding by the EU Research Training Network PLATON, contract number HPRN-CT-2000-00153.
NV acknowledges support by the U.S. Department of Energy under Grant No. B341495 to the Center for Astrophysical Thermonuclear Flashes at the University of Chicago and from NASA grant NAG 5-9063.
\end{acknowledgements}

\end{document}